\documentclass{epl}
\usepackage{epsfig}
\usepackage{float}
\usepackage{wrapfig}
\usepackage{amsmath}

\newcounter{blah}
\newcommand{\lettereqn}{\setcounter{blah}{\value{equation}}%
\setcounter{equation}{0}%
\addtocounter{blah}{1}%
\renewcommand{\theequation}{\mbox{\arabic{blah}\alph{equation}}}%
}
\newcommand{\backeqn}{\setcounter{equation}{\value{blah}}%
\renewcommand{\theequation}{\arabic{equation}}}

\title{Magnetic properties of the Anderson model: a local moment approach}
\shorttitle{Magnetic properties of the Anderson model}
\author{D. E. Logan\inst{1} \and N. L. Dickens\inst{1}}
\institute{
  \inst{1} University of Oxford,  Physical and \\
Theoretical Chemistry, South Parks Road,\\
Oxford OX1 3QZ, UK.
}

\begin{document}

\maketitle

\begin{abstract}
We develop a local moment approach to static properties of the symmetric Anderson model in the presence of a magnetic field, focussing in particular on the strong coupling Kondo regime.  The approach is innately simple and physically transparent; but is found to give good agreement, for essentially all field strengths, with exact results for the Wilson ratio, impurity magnetization, spin susceptibility and related properties.
\end{abstract}





Theoretical study of Anderson impurity models (AIM) \cite{ref1:anderson} has long formed a cornerstone of research in strongly correlated electron systems \cite{ref2:sc-es}. Its pursuit has recently acquired added impetus both from the experimental discovery that direct, tunable mesoscopic realisations of AIMs are provided by nanostructure devices such as quantum dots \cite{ref3:qdots}, or surface atoms probed by scanning tunneling microscopy \cite{ref4:stm}; as well as from the advent of dynamical mean-field theory for correlated lattice-based fermions \cite{ref5:CES,ref6:RMP}, within which models such as the Hubbard or periodic Anderson lattices map onto an effective, self-consistent AIM.
There is however a disparity between progress in theoretical descriptions of static (thermodynamic and related) properties of AIMs, and dynamical properties such as single-particle excitations.  
The former are generally well understood \cite{ref2:sc-es}, at least for the conventional (metallic) AIM \cite{ref1:anderson}, using a variety of powerful techniques that often yield exact results but unfortunately do not in general permit ready access to dynamics.  
Extant theories for the latter by contrast, which in principle should also recover static properties as a limiting case and are approximate by necessity, suffer from well known qualitative limitations \cite{ref2:sc-es}.  
And while numerical approaches, notably Wilson's numerical renormalization group (NRG) \cite{ref7:NRG}, circumvent most beautifully this impasse and provide benchmark numerical results (see {\em e.g.} \cite{ref2:sc-es}), the need for new, approximate theoretical methods is not thereby vitiated.

We have recently initiated development of one such --- the local moment approach (LMA) \cite{ref8:LMA,ref9:sg-LMA} --- a non-perturbative many-body method in which the notion of local moments \cite{ref1:anderson} is introduced explicitly and self-consistently from the outset.  
The underlying approach is simple and physically transparent, but not without virtue: for example it handles single-particle dynamics on all energy scales, while recovering correctly Fermi liquid behaviour at low energies; capturing in particular the spin-fluctuation physics charactristic of the strong coupling Kondo regime, where it leads to good quantitative agreement with NRG results for {\em e.g.} the universal single-particle scaling spectrum (see fig.~10 of \cite{ref10:LMA+NRG}).
The LMA is not moreover specific to the Fermi liquid behaviour ubiquitous in the conventional AIM.  
It can also handle non-Fermi liquid phases such as arise in the soft-gap AIM, to which the approach has also recently been applied \cite{ref9:sg-LMA} and compared successfully to NRG results in \cite{ref10:LMA+NRG}.

Static properties resultant from the LMA have not however been considered hitherto.  
That is the purpose of the present paper: to consider the LMA for the conventional symmetric AIM, generalised to include the effect of an applied magnetic field, in order to determine static properties and their field dependence, notably the excess impurity magnetization (and hence linear conductance of a quantum dot (see below)), the spin susceptibility and Wilson ratio.  
Our primary emphasis is naturally on the strong coupling Kondo regime, where we make comparison to and and find striking agreement with exact results provided by the Bethe ansatz \cite{ref11:BA}.  
This provides an acid test for any approximate theory, and since the LMA is confined neither to statics nor to integrable models, its success in describing quantum impurity models for which exact results {\em are} available provides important evidence for the strength of this new approach in generic situations where exact results will not typically be possible.

The AIM Hamiltonian is given in standard notation by

\begin{equation}
\label{eq:ham}
\hat{H} = \sum_{\bm{k},\sigma} \epsilon_{\bm{k}} \hat{n}_{\bm{k}\sigma} + \sum_\sigma \Big{(} \epsilon_{i\sigma} + \frac{U}{2} \hat{n}_{i -\sigma} \Big{)} \hat{n}_{i \sigma} + \sum_{\bm{k},\sigma} V_{i\bm{k}} \Big{(} c_{i \sigma}^{\dagger} c_{\bm{k}\sigma} + {\rm h.c.} \Big{)} .
\end{equation}

\noindent The first term refers to the non-interacting host; and the second to the impurity with on-site interaction $U$.  $\epsilon_{i\sigma} = \epsilon_{i} - \sigma h$ includes the (local) Zeeman coupling to the external field $H$, with $h = \frac{1}{2}g\mu_{{\rm B}}H$ and $\sigma = +/-$ for $\uparrow/\downarrow$-spin electrons.
For the symmetric AIM considered, $\epsilon_i = -\frac{U}{2}$; and by particle-hole (p-h) symmetry the impurity charge $n_i = \sum\nolimits_\sigma < \hat{n}_{i \sigma} > = 1~\forall~h$.
The final term in eq.~(\ref{eq:ham}) is the host-impurity coupling.
The total impurity Green function $G(\omega; h) = \frac{1}{2} \sum\limits_\sigma G_\sigma (\omega, h)$, with $G_\sigma(\omega; h) = G_\sigma^{{\rm R}}(\omega; h) - i~{\rm sgn}(\omega) \pi D_\sigma(\omega; h)$ given by

\begin{equation}
\label{eq:G}
G_\sigma(\omega; h) = \Big{[} \omega^+ - \Delta(\omega) + \sigma h - \tilde{\Sigma}_{\sigma} (\omega; h) \Big{]}^{-1}
\end{equation}

\noindent where $\omega^+ = \omega + i 0^+{\rm sgn}(\omega)$.
Here $\Delta(\omega) = \Delta_{{\rm R}} (\omega) - i~{\rm sgn}(\omega)\Delta_{{\rm I}}(\omega)$ is the host-impurity hybridization, with $\Delta_{{\rm I}}(\omega) = \pi \sum\nolimits_{\bm{k}} V_{i \bm{k}}^2 \delta(\omega - \epsilon_{\bm{k}})$; in practice we consider explicitly the wide-band AIM with $\Delta_{{\rm I}}(\omega) = \Delta_0~\forall~\omega$ (and $\Delta_{{\rm R}}(\omega) = 0$).
$\tilde{\Sigma}_\sigma (\omega; h) = \tilde{\Sigma}_\sigma^{{\rm R}} (\omega; h) - i~{\rm sgn}(\omega) \tilde{\Sigma}_\sigma^{{\rm I}}(\omega; h)$ denotes the self-energy (excluding the trivial Hartree contribution); by p-h symmetry $\tilde{\Sigma}_\sigma (\omega; h) = -\tilde{\Sigma}_{-\sigma} (-\omega; h)$ and $D_\sigma(\omega; h) = D_{-\sigma}(-\omega; h)$, such that $D_\sigma (0;h) \equiv D(0;h)$ at the Fermi level $\omega = 0$.
The h-dependent (but $\sigma$-independent) quasiparticle weight is defined by $z(h) = \big[ 1 - \big( \partial \tilde{\Sigma}_\sigma^{{\rm R}}(\omega; h) / \partial\omega \big)_{\omega = 0} \big]^{-1}$.

While $\tilde{\Sigma}_\sigma(\omega;h)$ in its entirety naturally determines spectral dynamics, its $\omega = 0$ limit enables the $h$-dependence of a range of important static properties to be deduced.  The single-particle spectrum at the Fermi level is given by $\pi \Delta_0 D(0;h) = \big[ 1 + \big( [h - \sigma \tilde{\Sigma}_\sigma^{{\rm R}} (0;h) ] / \Delta_0 \big)^2 \big] ^{-1}$; it determines the ($T = 0$) linear conductance of a quantum dot modelled by an AIM, via \cite{ref12:meir} $G_{\rm c}(h) = (e^2 / \pi \hbar) \pi \Delta_0 D(0;h)$ (for equal dot-lead couplings).  Likewise the (excess) impurity magnetization $M_{{\rm i}}(h)$ follows from the Friedel sum rule as \cite{ref2:sc-es}

\begin{equation}
\label{eq:genmag}
M_{{\rm i}}(h) = \frac{g \mu_{{\rm B}}}{\pi} \tan^{-1} \Big(\big[ h - \sigma \tilde{\Sigma}_\sigma^{{\rm R}} (0;h) \big] / \Delta_0 \Big)
\end{equation}

\noindent (for either spin $\sigma$), with the corresponding susceptibility $\chi_{{\rm i}}(h) = \partial M_{{\rm i}} / \partial H$ given by

\begin{equation}
\label{eq:genchi}
\chi_{{\rm i}}(h) = \frac{_1}{^2} (g \mu_{{\rm B}} )^2 D(0;h) \big[ 1 - \sigma \big( \partial \tilde{\Sigma}_\sigma^{{\rm R}} (0;h) / \partial h \big) \big].
\end{equation}

\noindent And the Wilson ratio, $R_{{\rm W}}(h) = \alpha \chi_{{\rm i}}(h) / \gamma_{{\rm i}}(h)$ with $\gamma_{{\rm i}}(h)$ the linear specific heat coefficient ($\alpha = [ 2 \pi k_{{\rm B}} ]^2 / [ 3 (g\mu_{{\rm B}})^2]$), is given by 
\begin{equation}
\label{eq:genrw}
R_{{\rm W}} (h) = z(h) \big[ 1 - \sigma \big( \partial \tilde{\Sigma}_\sigma^{{\rm R}} (0;h) / \partial h \big) \big] .
\end{equation}

\begin{wrapfigure}{l}{40mm}
\centering\epsfig{file=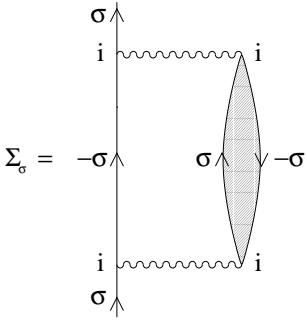,width=40mm,angle=0}
\protect\caption{LMA $\Sigma_\sigma(\omega)$, see text.  Wavy lines denote $U$.}
\label{fig:diag}
\end{wrapfigure}

Within the LMA \cite{ref8:LMA,ref9:sg-LMA}, $\tilde{\Sigma}_\sigma^{{\rm R}} (\omega;h) = - \frac{\sigma}{2}U | \mu(h) | + \Sigma_\sigma(\omega;h)$ is separated into a purely static Fock contribution (with local moment $|\mu(h)|$) that alone would survive at the crude mean-field (MF) level of unrestricted Hartree-Fock; plus a dynamical contribution $\Sigma_\sigma(\omega;h)$.
The latter includes in particular a non-perturbative class of diagrams (fig.~\ref{fig:diag}) that embody dynamical coupling of single-particle excitations to low-energy transverse spin fluctuations, and hence capture the spin-flip scattering essential to describe the strong coupling (SC) or Kondo regime for $\tilde{U} = U/ \pi \Delta_0 \gg 1$.
These are expressed in terms of MF propagators (solid lines in fig.~\ref{fig:diag}),

\begin{equation}
\label{eq:lma_G}
{\cal G}_\sigma (\omega;h) = \Big[ \omega^+ - \Delta(\omega) + \sigma \big( \frac{_1}{^2} U |\mu(h)| + h \big) \Big]^{-1}
\end{equation}

\noindent  and for $\omega = 0$ in particular $\Sigma_\uparrow(0;h)$ $\big( \equiv \Sigma_\uparrow^{{\rm R}}(0;h) \big)$ is given by

\begin{equation}
\label{eq:lma_sig}
\Sigma_\uparrow^{{\rm R}}(0;h) = U^2 \int\limits_{-\infty}^{\infty} \frac{{\rm d}\omega_1}{\pi}{\rm Im}\Pi^{+-}(\omega_1) \Big[ \theta(\omega_1){\rm Re}{\cal G}_\downarrow^-(\omega_1; h) + \theta(-\omega_1) {\rm Re} {\cal G}_\downarrow^+(\omega_1; h) \Big].
\end{equation}

\noindent The transverse spin polarization propagator $\Pi^{+-}(\omega)$ (shown hatched in fig.~\ref{fig:diag}) is given at the simplest level by an RPA-like particle-hole ladder sum in the transverse spin channel; ${\cal G}^{(\pm)}_\sigma (\omega)$ denote the one-sided Hilbert transforms of ${\cal G}_\sigma(\omega;h)$.

{\em LMA : $h=0$.} We first review briefly the LMA for $h=0$, an understanding of which underpins the general case $h \ne 0$; full details are given in \cite{ref8:LMA}.  Here the key idea is symmetry restoration, reflected in $\tilde{\Sigma}_\uparrow^{\rm R} (0;0) =\tilde{\Sigma}_\downarrow^{\rm R} (0;0)$; {\em i.e.} (using p-h symmetry)

\begin{equation}
\label{eq:sigma_sym}
\tilde{\Sigma}_\uparrow^{\rm R} (0;0) = - \frac{1}{2} U |\mu(0)| + \Sigma_\uparrow^{\rm R}(0;0) = 0.
\end{equation}

\noindent Imposition of eq.~(\ref{eq:sigma_sym}) as a self-consistency condition, achieved in practice for given $\tilde{U}$ by varying the local moment $|\mu(0)|$ from its MF value, preserves the $U$-independent pinning of the Fermi level spectrum ($\pi\Delta_0 D(0;0) =1$); and in turn leads to Fermi liquid behaviour at low-energies.
Most importantly, it introduces naturally a low-energy spin-flip scale $\omega_{\rm m}^0$ that sets the timescale for restoration of the broken symmetry endemic at MF level, and is manifest in a strong resonance centred on $\omega = \omega_{\rm m}^0$ in ${\rm Im}\Pi^{+-}(\omega)$.  This is the Kondo scale.  Its form in SC is readily deduced via eq.~(\ref{eq:lma_sig}), which for $\tilde{U} \gg 1$ has the asymptotic form \cite{ref8:LMA}

\begin{equation}
\label{eq:sc_sig}
\Sigma_\uparrow^{\rm R}(0;0) = \frac{4 \Delta_0}{\pi} \ln \left[ \frac{U}{\omega_{\rm m}^0}\right].
\end{equation}

\noindent Combining with eq.~(\ref{eq:sigma_sym}) (using $|\mu| \rightarrow 1$ in SC) gives the SC Kondo scale for $h=0$, viz $\omega_{\rm m}^0 \sim U \exp \big( -\pi U / 8 \Delta_0 \big)$ where the exponent is exact (the prefactor, in effect a uv-cutoff, is of course approximate).  
The leading low-$\omega$ behaviour of $\tilde{\Sigma}_\sigma^{\rm R}(\omega; 0)$ is also readily obtained \cite{ref8:LMA} in SC: $\big( \partial \tilde{\Sigma}_\sigma^{\rm R} (\omega;0) / \partial \omega \big) _{\omega = 0} \sim -4 \Delta_0 / \pi \omega_{\rm m}^0$, producing a quasiparticle weight $z(0) \sim \pi \omega_{\rm m}^0 / 4 \Delta_0$ that is naturally proportional to the Kondo scale.

{\em LMA: $h>0$.}  For $h=0$, two degenerate MF states arise \cite{ref8:LMA}.  The degeneracy is naturally removed for $h \ne 0$, where one or other state is picked out according to ${\rm sgn} (h)$.
The appropriate MF propagators then determine the many-body LMA self-energies $\tilde{\Sigma}_\sigma$, which satisfy $\tilde{\Sigma}_\sigma (\omega;h) = \tilde{\Sigma}_{-\sigma} (\omega;-h)$ as well as p-h symmetry (and hence {\em e.g.} $M_{\rm i}(h) = -M_{\rm i}(-h)$).
For convenience in what follows we consider explicitly $h > 0$.
Our primary aim is to consider the SC Kondo limit of the AIM.
To this end we thus consider finite $h / \Delta_0 z(0)$ in the SC limit where $z(0) \rightarrow 0$.

Since $G_\sigma = \big[ {\cal G}_\sigma^{-1} - \Sigma_\sigma\big]^{-1}$, and the LMA $\Sigma_\sigma \equiv \Sigma_\sigma \big[{\cal G}_\sigma \big]$ is a functional of the MF propagators given by eq.~(\ref{eq:lma_G}), it follows generally that the h-dependence is embodied fully in $x(h) = \frac{U}{2} |\mu(h)| + h$ (with $|\mu(0)|$ such that eq.~(\ref{eq:sigma_sym}) is satisfied for $h=0$).
The LMA $\tilde{\Sigma}_\sigma(\omega;h)$ for $h > 0$ is thus formally equivalent to that for $h=0$, but with $x(0) \rightarrow x(h) = x(0) + \big[ \frac{U}{2}\delta |\mu (h)| + h \big]$ (where $\delta |\mu (h)| = |\mu (h)| - |\mu (0)|$ is calculated in practice at MF level).
In consequence, $\Sigma_\sigma^{\rm R}(0;h)$ and $\big( \partial \tilde{\Sigma}_\sigma^{\rm R}(\omega; h) / \partial \omega \big)_{\omega = 0}$ in SC have idential forms as for $h=0$, but with $\omega_{\rm m}^0 = \omega_{\rm m}(h=0)$ replaced by $\omega_{\rm m}(h) \equiv \omega_{\rm m}(x(h))$ where $\omega_{\rm m}(h)$ is again the spin-flip scale characteristic of ${\rm Im}\Pi^{+-}(\omega)$; {\em i.e.} $\Sigma_\downarrow^{\rm R}(0;h) \sim \frac{4 \Delta_0}{\pi} \ln\big[ U / \omega_{\rm m}(h) \big]$, and $\big( \partial \Sigma_\sigma^{\rm R} (\omega;h) / \partial \omega \big)_0 \sim -4\Delta_0 / \pi \omega_{\rm m}(h)$ such that $z(h) / z(0) = \omega_{\rm m}(h) / \omega_{\rm m}^0$.  Hence, using eq.~(\ref{eq:sigma_sym}) for $h=0$,

\lettereqn
\begin{equation}
\label{eq:sig_h}
\tilde{\Sigma}_\uparrow^{\rm R} (0;h) = - \frac{4 \Delta_0}{\pi} \ln \left[\frac{z(h)}{z(0)}\right].
\end{equation}

\noindent For finite $h / \Delta_0 z(0)$, the $-\frac{U}{2}\delta | \mu(h) |$ contribution to $\tilde{\Sigma}_\uparrow^{\rm R}(0;h)$ is readily shown to be exponentially small in SC and is thus neglected in eq.~(\ref{eq:sig_h}).  
By the same token $\omega_{\rm m}(h) = \omega_{\rm m}(x(h))$ satisfies ${\rm d}\omega_{\rm m} / {\rm d}h = \partial \omega_{\rm m} / \partial x$; with $\partial \omega_{\rm m} / \partial x = 2$ in SC known from the LMA for $h=0$ \cite{ref8:LMA}.
Hence $\omega_{\rm m}(h) = \omega_{\rm m}^0 + 2h$, {\em i.e.}

\begin{equation}
\label{eq:z_h}
\frac{z(h)}{z(0)} = 1 + \frac{\pi}{2} \frac{h}{\Delta_0 z(0)} = 1 + \frac{\pi^2}{8} \frac{h}{k_{\rm B}T_0}.
\end{equation}
\backeqn


\noindent Here we have also introduced the scale $T_0$ defined by $k_{\rm B}T_0 = \pi \Delta_0 z(0) / 4$, in terms of which the exact linear susceptibility  in the SC Kondo limit is \cite{ref2:sc-es,ref11:BA} $\chi_{\rm i}(0) = (g \mu_{\rm B})^2 / 4k_{\rm B}T_0$.

Eqs.~(10) are the basic LMA equations appropriate to the SC Kondo limit of the AIM.  The Wilson ratio follows directly from eqs.~(\ref{eq:genrw}) and (10).
It is given by $R_{\rm W}(h) = 2$ for all finite $\tilde{h} = h / k_{\rm B}T_0$ and $k_{\rm B}T_0 \rightarrow 0$.  This is the exact result for the Kondo limit, including its field (in)dependence \cite{ref13:wiegmann}.
It is not however recovered by conventional approaches such as the slave boson mean-field approximation or $\frac{1}{N}$-expansions to leading order (with $N$ the impurity degeneracy), which give \cite{ref2:sc-es} $R_{\rm W} (0) = 1$ corresponding to non-interacting quasiparticles.

The magnetization in SC likewise follows from eqs.~(10) with (\ref{eq:genmag}) (where the `bare' h may be neglected since $h = \tilde{h} k_{\rm B}T_0 \rightarrow 0$); namely (for $h > 0$)

\begin{equation}
\label{eq:scmag}
M_{\rm i}(h) = \frac{g \mu_{\rm B}}{\pi} \tan^{-1} \left[ \frac{4}{\pi} \ln \left(1 + \frac{\pi^2}{8}\frac{h}{k_{\rm B}T_0} \right)\right].
\end{equation}

\noindent This is exact asymptotically for both weak and strong fields $\tilde{h} = h / k_{\rm B} T_0$.  For $\tilde{h} \ll 1$ it yields $M_{\rm i}(h) \sim \chi_{\rm i}(0) H$ with $\chi_{\rm i}(0) = (g\mu_{\rm B})^2 / 4k_{\rm B}T_0$, thus capturing the exact $\chi_{\rm i}(0)$.
For $\tilde{h} \gg 1$, eq.~(\ref{eq:scmag}) gives

\begin{equation}
\label{eq:bigh_mag}
M_{\rm i}(h) = \frac{g\mu_{\rm B}}{2} \left( 1 - \frac{1}{2\ln \left[ \frac{\pi^2}{8}\tilde{h}\right]} + {\rm O}\big( [\ln \tilde{h}]^{-3} \big) \right).
\end{equation} 

\begin{figure}[H]
\centering\epsfig{file=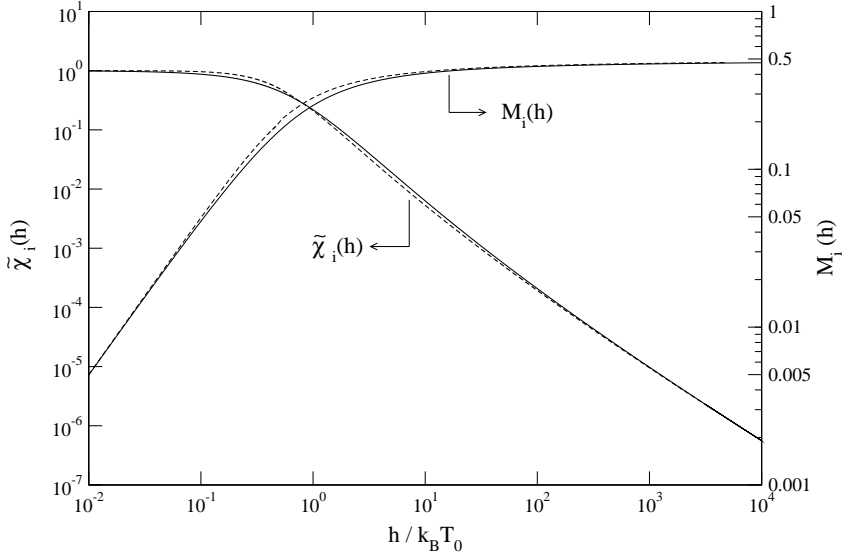,width=90mm,angle=270}
\vskip-5mm
\protect\caption{$M_{\rm i}(h)/g\mu_{\rm B}$ and ${\tilde{\chi}_{\rm i}}(h) = [4k_{\rm B} T_0 / (g \mu_{\rm B})^2]\chi_{\rm i}(h)$ vs $\tilde{h} = h / k_{\rm B}T_0$ in SC Kondo limit. Solid line: LMA. Dashed line: exact Bethe ansatz \cite{ref11:BA} for Kondo model.}
\label{fig:lma_v_ba}
\end{figure}

\noindent It thus recovers the slow logarithmic approach to saturation, although only the leading such correction is obtained, the exact $M_{\rm i}(h)$ being given by \cite{ref11:BA} $2M_{\rm i} / g\mu_{\rm B} = 1-[2\ln(\sqrt{\pi e}\tilde{h})]^{-1} + {\rm O}( \ln \ln \tilde{h} / [\ln \tilde{h}]^2)$.  
Fig.~\ref{fig:lma_v_ba} shows the full $\tilde{h}$-dependence of $M_{\rm i}(h)$ compared to the exact Bethe ansatz result for the Kondo model \cite{ref11:BA}. The agreement is very good throughout; the deviation is at most $\sim 10 \%$, occuring for fields $\tilde{h} \sim {\rm O}(1)$.  
Hence the field-dependence of the linear conductance, given (using eq.~(\ref{eq:genmag})) by $G_{\rm c}(h) = (e^2 / \pi \hbar) \cos^2[\pi M_{\rm i}(h) / g \mu_{\rm B}]$, is in turn well captured by the LMA, recovering the exact limits $(\pi\hbar/e^2) G_{\rm c}(h) \sim 1 - [\pi \tilde{h} / 2 ]^2$ for $\tilde{h} \ll 1$ and $\sim [\frac{4}{\pi}\ln(\tilde{h})]^{-2}$ for $\tilde{h} \gg 1$. 
Similar comments apply to the susceptibility $\chi_{\rm i} (h)$ which is also shown in fig.~\ref{fig:lma_v_ba}, and is likewise exact asymptotically in the weak and strong field limits.  Finally, since the LMA recovers correctly $R_{\rm W}(h) = 2$ for all $\tilde{h}$, the $\tilde{h}$-dependence of the specific heat coefficient $\gamma_{\rm i}(h) \propto \chi_{\rm i}(h)$ is precisely that of $\chi_{\rm i}(h)$.

\begin{figure}[H]
\centering\epsfig{file=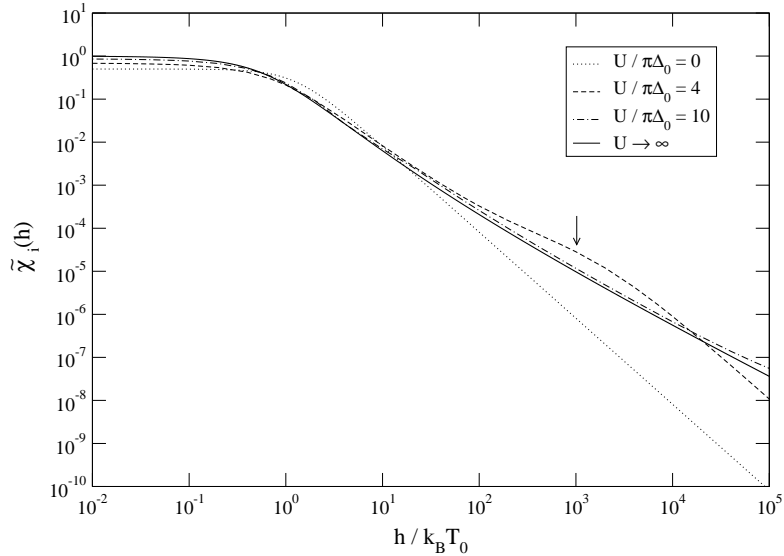,width=90mm,angle=270}
\vskip-5mm
\protect\caption{$\tilde{\chi}_{\rm i}(h) = [4k_{\rm B}T_0/(g\mu_{\rm B})^2]\chi_{\rm i}(h)$ vs $h/k_{\rm B}T_0$ for: the Kondo limit (solid line), $\tilde{U}=10$ (point-dash), 4 (dashed), and 0 (dotted).  For $\tilde{U}=4$, $h = U/2$ is marked by a vertical arrow.}
\label{fig:chi_v_U}
\end{figure}

Having discussed the SC Kondo limit, we now turn briefly to LMA results for finite-$\tilde{U}$, focusing on the susceptibility $\chi_{\rm i}(h) \equiv \chi_{\rm i}(h; \tilde{U})$ (eq.~(\ref{eq:genchi})).  Fig.~\ref{fig:chi_v_U} shows the $\tilde{h} = h/k_{\rm B}T_0$ dependence of $\tilde{\chi}_{\rm i}(h;\tilde{U}) = [4k_{\rm B}T_0 / (g \mu_{\rm B})^2] \chi_{\rm i}(h; \tilde{U})$ for $\tilde{U}$ = 10, 4 and 0; as above $k_{\rm B}T_0 = \pi \Delta_0 z(0) / 4$, with $z(0)$ the $\tilde{U}$-dependent, zero-field quasiparticle weight.  
For the trivial non-interacting limit (where $z(0) = 1$), $\tilde{\chi}_{\rm i}(h;0) = \frac{_1}{^2}[1+(\pi \tilde{h} / 4)^2]^{-1}$ and $\frac{\pi}{4}\tilde{h} \equiv \frac{h}{\Delta_0}$.
Note first that for $h=0$, $\tilde{\chi}_{\rm i}(0;\tilde{U}) = \frac{_1}{^2}R_{\rm W}(0)$ follows generally from eqs.~(\ref{eq:genchi}), (\ref{eq:genrw}) using $\pi \Delta_0 D(0,0) = 1$.  The LMA recovers correctly $R_{\rm W}(0) = 2$ in the Kondo limit, but does not yield the exact exponential approach (in $\tilde U$) to it, giving instead algebraic behaviour ($R_{\rm W}(0) \sim 2 - {\rm O}(1/\tilde{U}))$; $\tilde{\chi}_{\rm i}(0; \tilde{U})$ is thus underestimated somewhat by the LMA.
Nonetheless for $\tilde{U} = 10$, and over essentially the entire $\tilde{h}$-range shown in fig.~\ref{fig:chi_v_U}, $\tilde{\chi}_{\rm i}$ is very close to its universal SC Kondo limit (fig.~\ref{fig:chi_v_U}, solid line), exhibiting in particular the high-field behaviour $\tilde{\chi}_{\rm i} \sim [\tilde{h} \ln ^2 (\tilde{h})]^{-1}$ for 
$\tilde{h} {{_>}\atop{^\sim}} 10^2$.

For sufficiently large fields at any finite $\tilde{U}$ however, $\tilde{\chi}_{\rm i}(h;\tilde{U})$ should exhibit a crossover to behaviour charactristic of the free-orbital regime.  That the LMA correctly captures such behaviour is illustrated in fig.~\ref{fig:chi_v_U} for $\tilde{U} = 4$, where $h=\frac{U}{2}$ is marked by an arrow.  $\tilde{\chi}_{\rm i}(h;\tilde{U})$ is seen to cross over from Kondo-like behaviour and for $h \gg \frac{U}{2}$ becomes proportional (parallel in fig.~\ref{fig:chi_v_U}) to the non-interacting $\tilde{\chi}_{\rm i}(h;0)$ (shown dotted in fig.~\ref{fig:chi_v_U}).  
This is indeed the free-orbital regime: here the susceptibility itself, $\chi_{\rm i}(h;\tilde{U})$, is in practice $\tilde{U}$-independent and coincides with its $\tilde{U} = 0$ limit $\chi_{\rm i}(h;0) \sim [\Delta_0/h]^2$ (such that $\tilde{\chi}_{\rm i}(h;\tilde{U}) \propto z(0)[\Delta_0/h]^2$ for fields $h \gg \frac{U}{2}$).

In summary, we have developed a local moment approach to static magnetic properties of the Anderson model.  The approach encompasses all field and interaction strengths and, its undoubted simplicity notwithstanding, we find in particular for the strong coupling Kondo regime that it leads to good quantitative agreement with exact results obtained from sophisticated methods such as the Bethe ansatz \cite{ref2:sc-es,ref11:BA}.  However the LMA can handle in addition both non-integrable models and dynamical properties [6-8].  Its success in capturing static magnetic properties, particularly in the strong coupling regime, suggests its efficacy as a viable route to {\em e.g.} the field dependence of single-particle dynamics. We will discuss this important issue in a subsequent publication.

\acknowledgments{We are grateful to the EPSRC for financial support.}



\begin{thebibliography}{0}

\bibitem{ref1:anderson}
  \Name{Anderson P. W.}
  \REVIEW{Phys. Rev.}{124}{1961}{41}.

\bibitem{ref2:sc-es}
  \Name{Hewson A. C.}
  \Book{The Kondo Problem to Heavy Fermions}
  \Publ{CUP, Cambridge}
  \Year{1993}


\bibitem{ref3:qdots}
  \Name{Goldhaber-Gordon D. {\em et. al.}}
  \REVIEW{Nature}{391}{1998}{156}.

\bibitem{ref4:stm}
  \Name{Li J., Schneider W.-D., Berndt R. \and Delley B.}
  \REVIEW{Phys. Rev. Lett.}{80}{1998}{2893}.

\bibitem{ref5:CES}
  \Name{Vollhardt D.}
  \Book{Correlated Electron Systems}
  \Editor{Emery V. J.}
  \Vol{9}
  \Publ{World Scientific, Singapore}
  \Year{1993}
  \Page{57}.

\bibitem{ref6:RMP}
  \Name{Georges A., Kotliar G., Krauth W. \and Rozenberg M. J.}
  \REVIEW{Rev. Mod. Phys.}{68}{1996}{13}.

\bibitem{ref7:NRG}
  \Name{Krishnamurthy H. R., Wilkins J. W. and Wilson K. G.}
  \REVIEW{Phys. Rev. B}{21}{1980}{1003}; {\em ibid}, {\bf 21} (1980) 1044.

\bibitem{ref8:LMA}
  \Name{Logan D. E., Eastwood M. P. \and Tusch M. A.}
  \REVIEW{J. Phys. Condensed Matter}{10}{1998}{2673}.

\bibitem{ref9:sg-LMA}
  \Name{Logan D. E. \and Glossop M. T.}
  \REVIEW{J. Phys. Condensed Matter}{12}{2000}{985}.

\bibitem{ref10:LMA+NRG}
  \Name{Bulla  R., Glossop M. T., Logan D. E. \and Pruschke T.}
  \REVIEW{J. Phys. Condensed Matter}{12}{2000}{4899}.

\bibitem{ref11:BA}
  \Name{Andrei N., Furuya K. \and Lowenstein J.H.}
  \REVIEW{Rev. Mod. Phys.}{55}{1983}{331}.

\bibitem{ref12:meir}
  \Name{Meir Y., Wingreen N. S. \and Lee P.A.}
  \REVIEW{Phys. Rev. Lett.}{70}{1993}{2601}.

\bibitem{ref13:wiegmann}
  \Name{Wiegmann P. B. \and Finkelstein A. M.}
  \REVIEW{Sov. Phys. JETP}{48}{1978}{102}.






\end{thebibliography}
\end{document}